\documentclass[english,final,proceeding]{jcaa}
\usepackage{url}

\title{SOUND-STREAM II: Towards Real-time Gesture-controlled Articulatory Sound Synthesis}
\author{Pramit Saha\thanks{pramit@ece.ubc.ca}}
\author[]{Debasish Ray Mohapatra}
\author[]{Praneeth SV }
\author[]{Sidney Fels }
\affil[]{Human Communication Technologies Lab, Department of Electrical and Computer Engineering,\\ University of British Columbia}

\begin{document}

\twocolumn[

\maketitle

]

\saythanks

\section{Background and motivation}
Articulatory speech synthesis is utmost important for understanding the mechanism of human speech production. It encompasses the production of speech sounds using an artificial vocal tract model and simulating the movements of the speech articulators like tongue, lips, velum etc. However, despite having considerable significance in research and learning purposes, there is a dearth of intuitive user interfaces to effectively control the articulatory parameters based on simultaneous variation of speech articulators. 

This is because of the level of complexity of the vocal tract articulators, that participate in speech production
process. Vocal tract comprises of several organs, carefully controlled by the muscles. One of the key principles involved in articulatory synthesis lies in the simultaneous activation of these muscles to perform a multidimensional control of various parts of the vocal tract. Such movement occurs in an extremely interdependent manner, due to the intermingling of muscles. 

\section{Previous Works}

The prevalent user interfaces targeting such movements like Pink Trombone\cite{pt}, VT Demo\cite{mvt} etc utilize simple mouse-based kinematic control of midsagittal sliced tongue, lips, hard palate and velum. These controllers allow the user to manipulate individual parts of the tract - one at a time, to synthesize the vocal sounds. Furthermore, these changes occur in some predefined trajectories which are less intuitive and difficult to relate to the slider changes triggered by the user. There is a lack of user flexibility, since a user can achieve only one particular shape among a number of pre-defined tongue shapes, corresponding to changes in slider values. Furthermore, it essentially enables user to explore the effect of only one articulatory parameter or shape/deformation of one part of the tongue for production of vocal sound. The other parts of the same articulator or different articulators are assumed to be fixed.

Therefore, this kind of control becomes highly unrealistic when compared to the actual articulatory speech production process. In particular, the tongue is a highly deformable, muscular hydrostat organ with infinite degrees of freedom, equipped with eleven muscles (extrinsic and intrinsic) controlling its shape and position. Kinematic control of a handful of points on the tongue surface \cite{saha2018sound} ignores the practical biomechanical constraints and the neuro-muscular pathway behind speech. Hence, more research needs to be directed towards user-interface facilitating the control and manipulation of the tract contour including tongue. Besides, most of the interfaces allow mere independent controls of various parts, which means control of one part of the articulator do not reflect any changes in the other parts or do not provide any feedback to the user for the variations in other interrelated parts. However, in reality, our muscles and articulators are intimately interleaved and have biomechanical constraints, because of which, movement in one part of an articulator renders changes in other parts as well. To this end, we develop our SOUND STREAM Interface trying to develop a hand-manipulated force-based realistic tongue-control strategy for sound production.

\section{Overview of the proposed methodology}
We present an interface involving four degrees-of-freedom (DOF) mechanical control of a two dimensional, mid-sagittal tongue through a biomechanical toolkit called ArtiSynth \cite{stavness2011coupled} and a sound synthesis engine called JASS \cite{van2001jass} towards articulatory sound synthesis. As a demonstration of the project, the user will learn to produce a range of JASS vocal sounds, by varying the shape and position of the ArtiSynth tongue in 2D space through a set of four force-based sensors. In other words, the user will be able to physically play around with these four sensors, thereby virtually controlling the magnitude of four selected muscle excitations of the tongue to vary articulatory structure. This variation is computed in terms of `Area Functions' in ArtiSynth environment and communicated to the JASS based audio-synthesizer coupled with two-mass glottal excitation model to complete this end-to-end gesture-to-sound mapping. 
\begin{figure}[b]
  \centering
 {\includegraphics[width=8cm,height=5cm]{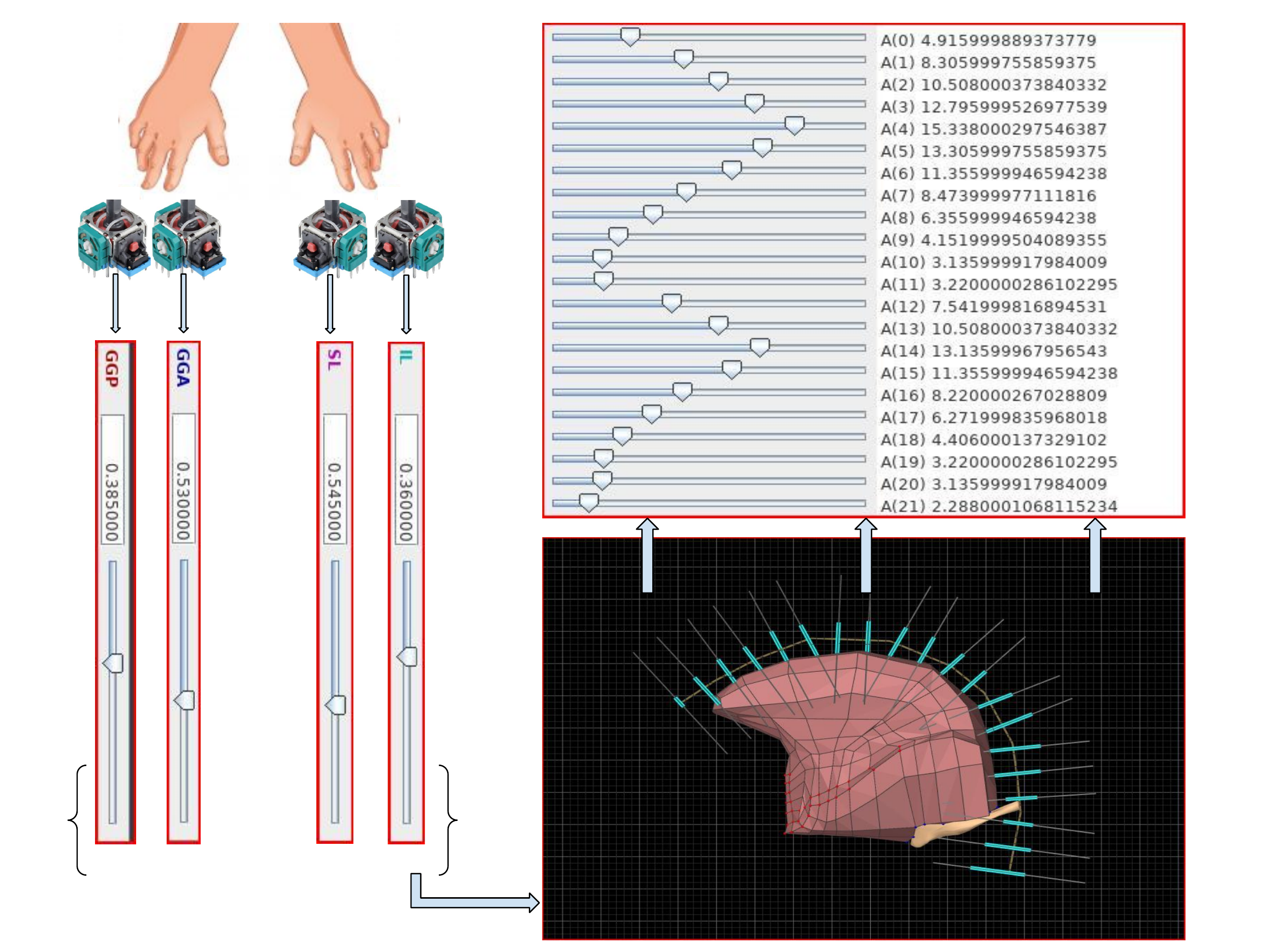}\label{fig:f2}}
  \caption{The proposed hand gesture-to-sound control pathway}
\end{figure}
\section{System Design}
Our hardware interface consists of four mini-joystick force sensors mounted on a fixed platform. These sensors are potentiometer based force sensitive resistors that measure the applied force. The finger pressure exerted on each of the joysticks result in changes of resistances
connected as a part of each voltage divider. Consequently, the analog input of Arduino microcontroller measures the output voltages, which are then translated to the tongue muscle excitations. The software interface consists of communication protocols between Arduino, ArtiSynth and JASS.

\section{Detailed Mechanism}
The proposed real-time gesture controlled sound synthesizer, through biomechanically-driven articulatory pathway, has three main phases (See Fig. 1), as discussed below:
\subsection{Gesture-to-muscle activations}
The first step is force-activated tongue muscle control, where we essentially replace the high-dimensional neural control of muscles by low dimensional hand gesture based tongue muscle manipulation. Here, we follow a simplistic force-to-muscle mapping strategy, where the tongue muscle activation ranging from 0 to 1, varies proportionally with the force exerted by the fingers.
\subsection{Muscle-to-movement}
We particularly select two intrinsic (Inferior and superior longitudinal) and two extrinsic muscle groups (Anterior and posterior genioglossus) to be controlled by the ambidextrous hand gestures. The longitudinal muscles are responsible for tongue retraction, making it short and thick. On the other hand, the genioglossus plays a major role in tongue protrusions and moving the tongue tip back and down. So variation of these muscle group excitations have significant effect on tongue shape and position. 

The established forward biomechanical pathway in ArtiSynth \cite{stavness2011coupled} allows conversion of muscle excitations to resultant movements. We utilized this to get tongue shape and position changes from the real-time variation of selected muscle activations. Next, we constructed a series of beams around the tongue, with 22 fixed markers set at regular intervals along the vocal tract surface, following \cite{Wang_2012}, as shown in Fig. 2. We further computed the distance between the tongue surface (varying with muscle activation changes) and these markers and derived the effective cross sectional area function for feeding it into the articulatory audio synthesizer.

\begin{figure}[]
  \centering
 {\includegraphics[width=9cm,height=3.5cm]{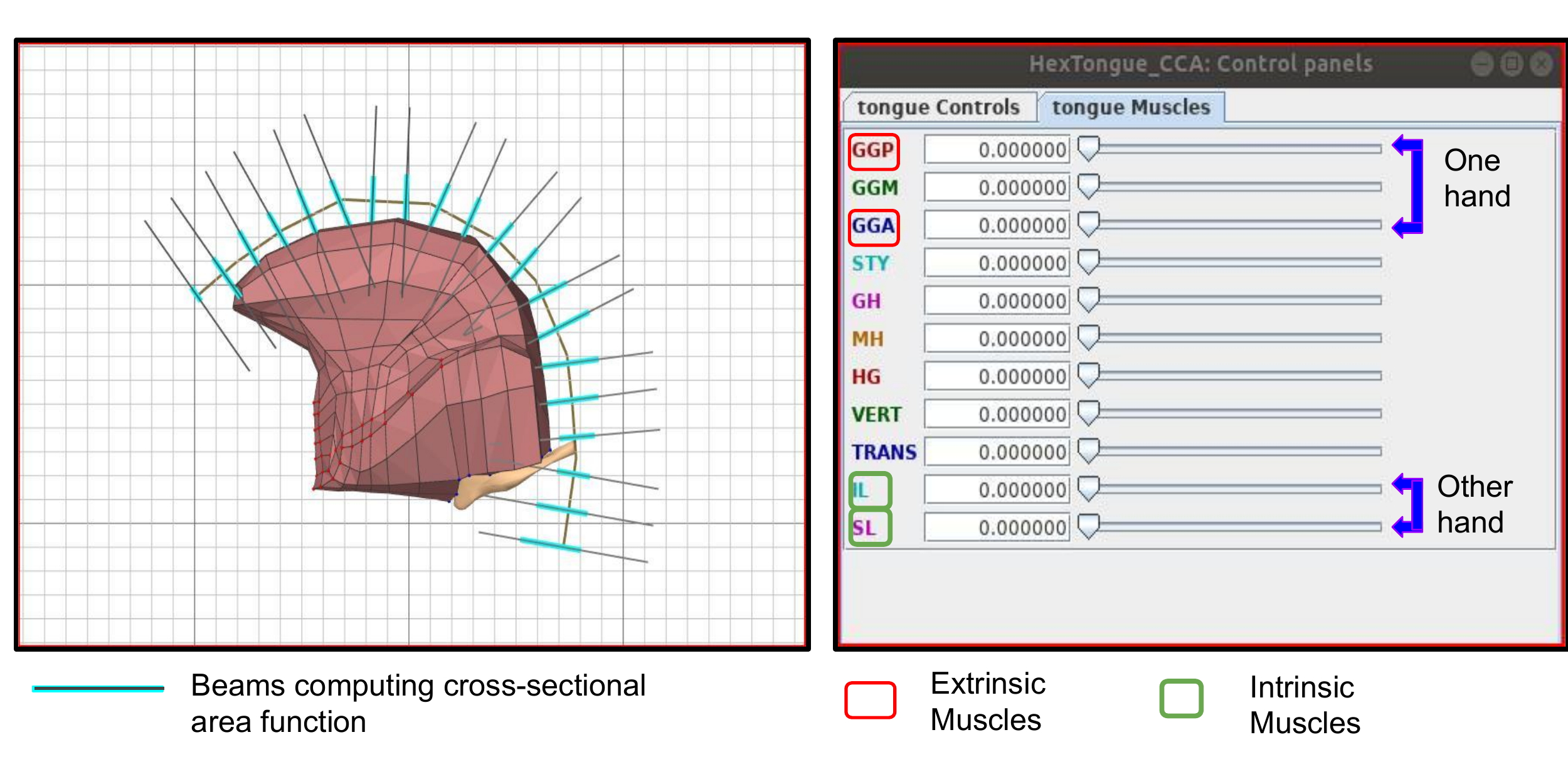}\label{fig:f1}}
  \caption{ArtiSynth Tongue Control}
\end{figure}
\subsection{Movement-to-Sound output}

The array of area functional values are sent in real-time to the Java Audio Synthesis System (JASS) which considers the vocal tract as an acoustic tube with its shape changing accordingly with the area functions. Glottal excitation pulse was generated according to the Rosenberg’s model and coupled to discretized acoustic equations in the vocal tract. The
acoustic wave propagation was simulated by numerically integrating the linearized 1D Navier-Stokes pressure-velocity PDE in time and space on a non-uniform grid. The synthesis mechanism involved excitations acting as source placed in the tube and sound propagation being simulated by approximating the pressure-velocity wave equations. 
\section{Discussions and Conclusions}
In this work, we explored a low-dimensional subspace of the high dimensional neuro-muscular control of tongue muscles, towards articulatory vocal sound synthesis. Using this interface, the user can use his/her fingers to play around with the muscle activations, to achieve real-time changes in tongue shape and position resuting in simultaneous variation of vocal sound. 

Therefore, this work offers an alternative pathway to the conventional kinematic approach of controlling vocal tract movements like \cite{saha2018sound,mvt,pt}. A qualitative pilot study on the proposed interface revealed that though the inexperienced users find it somewhat difficult to achieve target tongue movements quickly, they agree that this interface provides them with more variability of inputs and more intuitive understanding of the human voice synthesis. Hence, it can be concluded that the proposed force-activated vocal sound controller is indeed a step towards natural articulatory speech production.

\section*{Acknowledgments}
This work was funded by the Natural Sciences and Engineering Research Council (NSERC) of Canada and Canadian Institutes for Health Research (CIHR).

\bibliographystyle{unsrt}
\bibliography{jcaa}
\nocite{*}

\end{document}